\documentstyle[11pt,aaspp4,flushrt]{article}
\lefthead{DIAFERIO \& GELLER}
\righthead{INFALL REGIONS OF GALAXY CLUSTERS}
\begin{document}
\title{INFALL REGIONS OF GALAXY CLUSTERS}
\author{Antonaldo Diaferio\footnote{Present address:
Max-Planck-Institut f\"ur Astrophysik, Karl-Schwarzschild-Str. 1, 85740, Garching
bei M\"unchen, Germany} and Margaret J. Geller}
\affil{Harvard-Smithsonian Center for Astrophysics, 60 Garden Street Cambridge, MA 02138
\\ ApJ, in press. Scheduled for the June 1, 1997 issue, Vol. 481} 
\authoraddr{diaferio@mpa-garching.mpg.de}

\begin{abstract}
In hierarchical clustering, galaxy clusters accrete mass through the aggregation
of smaller systems. Thus, the velocity field of the infall regions of clusters contains 
significant random motion superimposed on radial infall. Because the purely spherical infall
model does not predict the amplitude of the velocity field correctly, 
methods estimating the cosmological density
parameter $\Omega_0$ based on this model yield unreliable biased results. In fact, 
the amplitude of the velocity field depends on local dynamics and only very weakly on 
the global properties of the universe.

We use $N$-body simulations of flat and open universes to show that
the amplitude of the velocity field of the infall regions of dark matter halos is a direct measure
of the escape velocity within these regions.
We can use this amplitude to estimate the mass of 
dark matter halos within a few megaparsecs from the halo center. In this region dynamical
equilibrium assumptions {\it do not} hold. The method yields a mass estimate with better than $30\%$
accuracy. If galaxies trace the velocity field of the infall
regions of clusters reliably, this method provides a straightforward 
way to estimate the amount of mass surrounding rich galaxy clusters from redshift
data alone.
\end{abstract}


\section{INTRODUCTION}

The linear theory of density perturbations shows that a spherically symmetric 
mass concentration in an expanding universe 
induces a radial peculiar velocity field in the surrounding region 
\begin{equation}
{{v_{\rm pec}(r)}\over {H_0r}} = -{{1}\over {3}}\Omega_0^{0.6} \delta(r) 
\label{1.1}
\end{equation}
where $\delta(r)$ is the average spherical mass overdensity within the radius $r$, and
$H_0$ and $\Omega_0$ are the Hubble constant\footnote{We use $h=0.5$ 
throughout, where $H_0=100h$ km s$^{-1}$ Mpc$^{-1}$.} 
and the cosmological density parameter at the present
time, respectively. We can estimate the galaxy number overdensity $\delta_g=b\delta$
where $b$ is the bias parameter. Thus, if we can measure $v_{\rm pec}$ 
we can estimate $\beta=\Omega_0^{0.6}/b$. 
Several authors have applied this method to the Local Supercluster and to the Virgo
cluster with mixed results (see e.g. the review by \cite{Davis83}; see also 
\cite{Strauss95}). Along with the uncertainties in the determination of the galaxy
number overdensity and the peculiar velocity themselves,  
external tidal shear can strongly affect the velocity field (e.g. \cite{Hoff86};
\cite{Eisen95}; \cite{Bond96}); thus we cannot obtain reliable results unless we sample 
the velocity field within the full three-dimensional region around the cluster 
(\cite{Vill86}).

In redshift space, spherical infall confines 
galaxies around clusters within caustics, surfaces with a characteristic ``trumpet'' 
shape (\cite{Kaiser87}). For these non-linear regions, we must 
replace the linear regime equation (\ref{1.1}) with the exact solution of the equation
of motion of shells in an overdense region within the expanding universe (e.g. \cite{Silk77}). However,
the dependence of the peculiar velocity on $\beta$ and $\delta_g$ is still approximately 
separable (\cite{Reg89}). 
Thus, provided we can determine the location of the caustics,
we can still estimate $\beta$ through the measurement of $\delta_g$ for a rich galaxy cluster 
(\cite{Reg89}). Recently, \cite{Reg96} suggested the application of
gravitational lensing (e.g. \cite{Tyson90}; \cite{Kais93}; \cite{Kais95}; \cite{Bonn95}) 
to determine the mass overdensity $\delta$ and therefore the
unbiased value of $\Omega_0$.

Standard inflationary cosmologies (e.g. \cite{Peacock96}) predict that the primordial density
field is a Gaussian random field. These initial conditions can lead to 
either top-down or bottom-up scenarios for the formation of cosmic structures. The hierarchical 
clustering in the bottom-up scenarios, with large systems forming by aggregation of
smaller ones, currently represents the most successful framework
of structure formation theories as it seems 
to be able to reproduce many, although not all, properties of the real universe. 
The top-down scenarios are far less succesfull 
(see e.g. \cite{Pad93} for a general discussion). 

The general formation process of clusters in hierarchical scenarios differs substantially
from the one described by the idealized spherical model; here clusters form through the infall
of smooth spherical shells onto an initial density peak (\cite{Gunn72}).
Despite its idealized nature, the spherical model appears to represent a reasonable description of 
the collapse of high peaks in a random Gaussian density field (\cite{Bern94}), and predicts 
density and velocity profiles of final systems in reasonably good agreement 
with $N$-body simulations of dark matter halo formation when the power spectrum 
has an effective spectral index $n\ge -1$ (\cite{Zar96}). 
These authors suggest that particle ranks in  
binding energy are conserved during the formation process. This conservation is responsible 
for the agreement (see also \cite{Hoffman88}; \cite{Quinn88}; \cite{Zar93}). 
In fact, for spectral indices $n\le -1$ the formation process is more violent (\cite{LyndenBell67}), 
energy ranks are not conserved, and the agreement breaks down. 

Thus, the agreement in energy space between the spherical infall model and 
hierarchical clustering for a limited range of spectral index, $n$, allows
correct predictions about the {\it final state} of the halo. However,   
the spherical infall model does not describe the {\it evolution} of the outer regions 
of systems in configuration space. $N$-body simulations of flat universes
(\cite{Haar92}; \cite{HaarWeyg93})
show that the velocity field of cluster surroundings is very different
from the predictions of spherical infall because (1) recent mergers increase the particle
kinetic energy, and (2) the presence of substructures makes the
velocity profile irregular. In other words, random motions obscure the infall information. 
If the same problem arises in real galaxy clusters, estimates of $\beta$ (or $\Omega_0$) based on 
spherical infall are systematic overestimates.

Here, we suggest a unifying explanation for the amplitude of the caustics. 
We use $N$-body simulations of flat and open universes to show that 
the escape velocity around dark matter halos determines the amplitude of the caustics.
Thus, the local dynamics of the halo, not the global
properties of the universe, dominates the amplitude of the velocity field around clusters.

We also show that under reasonable hypotheses about the density field 
outside the virialized region, the velocity field around a cluster provides an estimate of the 
mass enclosed within a few megaparsecs of the cluster center. This method is particularly 
interesting because, in this region, 
the equilibrium assumptions underlying usual mass estimation methods do not hold. 

In Sect. 2 we review the main equations of the spherical infall model. 
In Sect. 3 we derive the alternative expression for the amplitude of the caustics and we 
outline the mass estimation method.
In Sect. 4 we compare both the expression for the amplitude of the caustics 
and the spherical infall model with the results of $N$-body simulations. 
We also use these $N$-body simulations to test our new mass estimation method.   

\section{THE SPHERICAL INFALL MODEL}

We now review the main points of the infall model for a spherical perturbation in the 
case of open and flat universes. We then define the infall region and the caustic surfaces 
surrounding the perturbation. 

Consider a spherical perturbation in an expanding universe described by the
average overdensity profile $\delta_i(<r_i)$ within the physical radius $r_i$ at 
time $a_i\ll 1$, where $a$ is the cosmological scale factor, $a=1$ at the present time, 
and $\delta_i\ll 1$ for any radius $r_i$. The evolution of each shell in
the absence of shell crossings is described by the
equation of motion $d^2r/dt^2 = -GM/r^2 + \Lambda r/3$, where $M$ is the mass 
within $r$, $\Lambda$ is the cosmological constant, and $G$ is the gravitational constant. 
The first integral of the equation of motion is 
\begin{equation}
{{1}\over {2}}\left({dr}\over {dt}\right)^2 - {{GM}\over {r}} - {{\Lambda}\over {6}}r^2= E  
\label{2.1}
\end{equation}
where $E$ is the total energy of the shell.
In order to consider the collapse independent of the primordial density profile,
we can think in terms of the local scale factor $x=ra_i/r_i$. Note that at time $a_i$, $x=a_i$ for 
all the shells. However, at later times, if we assume that $\delta_i$ is a monotonically
decreasing function of $r_i$, the degeneracy between $x$ and $r_i$ disappears, and we
have a biunique correspondence between $x$ and $\delta_i$.
With this change of variable, equation (\ref{2.1}) becomes
\begin{equation}
\dot x^2- {{\alpha}\over {x}} -\Omega_{\Lambda0}x^2 = e 
\label{2.2}
\end{equation}
where the dot indicates a derivative with respect to $\tau=H_0t$, 
$\Omega_{\Lambda0}=\Lambda/3H_0^2$, and 
\begin{eqnarray}
\alpha(\delta_i) &=& \Omega_0(1+\delta_i)  
\label{2.3a} \\
e(\delta_i) &=& 1-\Omega_{\Lambda0} -
\Omega_0\left(1+{5\over 3}{{\delta_i}\over {a_i}}\right).   
\label{2.3b} 
\end{eqnarray}

In deriving equation (\ref{2.3b}) for the effective energy $e=(2E/H_0^2)(a_i/r_i)^2$, 
we set the initial conditions at time $a_i\sim 10^{-3}$, when linear theory in a matter-dominated
universe is a good description of the growth of perturbations. In other words,
we assume that at this epoch the decaying modes
of the perturbation have decayed away and the peculiar velocity field $v_i$ has reached the
amplitude predicted by linear theory, $v_i/H_ir_i= -\Omega_{0i}^{0.6}\delta_i/3$,
where $H_i$ is the Hubble constant and $\Omega_{0i}$ is the cosmological density
parameter  at time $a_i$ (see e.g. \cite{Lilje91}).
Thus, to first order in $\delta_i$, the peculiar velocity field contributes the
term $-2\Omega_0\Omega_{0i}^{0.6}\delta_i/3a_i$ to the effective total energy.
In equation (\ref{2.3b}) we have further assumed 
$\Omega_{0i}=\Omega_0/[\Omega_0 +(1-\Omega_0-\Omega_{\Lambda 0})
a_i + \Omega_{\Lambda 0}a_i^3]\simeq 1$, when $a_i\sim 10^{-3} \ll 1$.
 
When $\Omega_{\Lambda0}<0$, the shell always collapses. Otherwise, for collapse to occur, 
the effective energy $e$ must satisfy
\begin{equation}
 e < \cases{ -3\Omega_{\Lambda0}(\alpha/ 2\Omega_{\Lambda0})^{2/3} &
$\Omega_{\Lambda0}>0$ \cr
 0 & $\Omega_{\Lambda0}=0.$ \cr } 
\label {2.4}
\end{equation}
The smallest positive root (or the only positive root when $\Omega_{\Lambda0}<0$) of the 
cubic equation $\Omega_{\Lambda0}x^3 + ex+\alpha=0$ gives the maximum expansion scale factor 
$x_{\rm max}(\delta_i)$ of the shell with internal overdensity $\delta_i$.
Thus, we may write the collapse time
\begin{equation}
\tau_{\rm coll}(\delta_i) =2\int_0^{x_{\rm max}(\delta_i)} {{dy}\over {\dot x(y,\delta_i)}}.
\label{2.5}
\end{equation}
Finally, we can write the equation describing the evolution 
of $x(\delta_i,\tau)$ implicitly 
\begin{equation}
\tau =\cases{ \int_0^{x(\delta_i,\tau)} dy/ \dot
x(y,\delta_i) & $0\le\tau\le\tau_{\rm coll}/2$ \cr
\tau_{\rm coll}(\delta_i) - \int_0^{x(\delta_i,\tau)} dy /
\dot x(y,\delta_i) & $\tau_{\rm coll}/2<\tau\le \tau_{\rm coll}$. \cr} 
\label{2.6} 
\end{equation}

In terms of the local scale factor $x$, we next determine the caustics which depend 
only on the cosmological parameters and not on the overdensity profile. 
Consider the spherical perturbation within space $X=(x,\theta,\varphi)$, where $\theta$ and
$\varphi$ are the azimuthal and longitudinal angle, respectively.
In the following, 
we use the terms ``distance'' and ``velocity'' to indicate the quantities $x$ and $\dot x$, 
respectively. Recall that $x=ra_i/r_i$. Thus, we need
the initial average overdensity profile $\delta_i(<r_i)$ to transform space $X$
into configuration space $R=(r,\theta,\varphi)$, and to obtain the physical distance
$r$ and the physical velocity $v=dr/dt$.

Now, suppose we project the spherical perturbation within space $X$ onto the plane 
$\theta=\pi/2$. We can measure the projected distance $x_\perp$ from the origin, the
longitudinal angle $\varphi$ and the velocity component 
$\dot x_3$ along the axis $\theta=0$.  The relations between the new and the old variables are
\begin{equation}
x_\perp = x(\delta_i,\tau)\sin\theta, \ \
\dot x_3 = \dot x(x,\delta_i) \cos\theta, \ \
\varphi=\varphi. 
\label{2.7}
\end{equation}

Suppose we observe the spherical perturbation at a time,
$\tau$. From the equations $\tau_{\rm coll}(\delta_i^c)=\tau$ and
$\tau_{\rm coll}(\delta_i^m)=2\tau$ (eq. [\ref{2.5}])  we can determine the primordial overdensity 
$\delta_i^c$ of
the shell which has just collapsed and the $\delta_i^m$ of the shell which is at maximum
expansion at time $\tau$. Thus, the local scale factor
of the perturbation at turnaround is $x_{\rm ta}=x_{\rm max}(\delta_i^m)$. 
To determine the virial scale factor at time $\tau$ we need to consider the virial theorem (see
e.g.  \cite{Lahav91}).
We have $e = -\alpha/2x_v-2\Omega_{\Lambda0}x_v^2$ in virial 
equilibrium, and $e = -\alpha/x_m - \Omega_{\Lambda0}x_m^2$ for the {\it same} shell
at maximum expansion. Energy conservation implies 
\begin{equation}
 2\eta\psi^3 -(2+\eta)\psi + 1=0 
\label{2.8}
\end{equation}
where $\psi=x_v/x_m$ and $\eta = 2\Omega_{\Lambda0}x_m^3/\alpha$. Note that we can
also write $\eta=\Lambda/4\pi G \rho_m$, where $\rho_m$ is the mean density
of the perturbation within $r_m=r_ix_m/a_i$.  
The condition for collapse is $\eta<1$, and
we are interested in the solution $\psi\le 1$ of equation (\ref{2.8}) with $\eta$ a free parameter.
We have
\begin{equation}
\psi(\eta) = \cases { 2p^{1/3} \cos[\arccos(q/p)/3] & $\eta\le \eta_0$\cr
 (q+\Delta^{1/2})^{1/3} + (q-\Delta^{1/2})^{1/3} & $\eta_0<\eta<0$ \cr
1/2 & $\eta=0$\cr
2p^{1/3} \cos[\arccos(q/p)/3 + 4\pi/3] & $0< \eta<1$ \cr} 
\label{2.9}
\end{equation}
where $p=[(2+\eta)/6\eta]^{3/2}$, $q=-1/4\eta$, $\Delta = q^2-p^2$, and $\eta_0\simeq
-6.427$ is the only real solution of the cubic equation $\Delta=0$.

Now we can write the virial scale factor $x_{\rm vir}$ at time $\tau$, 
$x_{\rm vir}=\psi(\eta)x_{\rm max}(\delta_i^c)$ where
$\eta=2\Omega_{\Lambda0}x_{\rm max}^3(\delta_i^c)/\Omega_0(1+\delta_i^c)$. 
We define the infall region of the
perturbation at time $\tau$ as the region within the scale factor range
$(x_{\rm vir},x_{\rm ta})$. Note that $x_{\rm vir}$ and $x_{\rm ta}$ are the 
virial scale factor and the turnaround scale factor
of two {\it different} shells of the perturbation.

At fixed $x_\perp \in (x_{\rm vir},x_{\rm ta})$,
the velocity component $\dot x_3$ depends only on
$\theta$, $\dot x_3=\dot x[x_\perp/\sin\theta,\delta_i(\theta)]\cos\theta$
(see eqs. [\ref{2.7}] and [\ref{2.2}]), 
where $\delta_i(\theta)$ satisfies the implicit equation (see eq. [\ref{2.6}])
\begin{equation}
\tau = \tau_{\rm coll}[\delta_i(\theta)] -
\int_0^{x_\perp/\sin\theta}
{dy\over \dot x[y,\delta_i(\theta)]}.
\label{2.10}
\end{equation}
Thus, the extrema of $\dot x_3$ occur at
$\theta=\theta_{\rm max}(x_\perp,\tau)$, and $\theta=\pi-\theta_{\rm
max}(x_\perp,\tau)$ which satisfy the equation $d\dot x_3/d\theta=0$.
Note that $\theta_{\rm max}$ is almost independent of the cosmology. 

In the plane ($x_\perp,\dot x_3$) the perturbation fills only
the region within the curves
\begin{equation}
 \dot x_3^2(x_\perp,\tau) = \dot x^2
\{x[\delta_i(\theta_{\rm max}),\tau]\} \cos^2\theta_{\rm max} 
\label{2.11}
\end{equation}
which determine the caustics.
For any $x_\perp$, the maximum difference in the velocity component $\dot x_3$ is the
difference between the velocities on the caustics, $\Delta \dot x_3(x_\perp,\tau) = 
2 \vert \dot x_3(x_\perp,\tau)\vert$.

To switch from space $X$ to configuration space $R$, 
we must take the initial average overdensity profile $\delta_i(<r_i)$ into account. However,  
at any time, the physical radius is $r=xr_i/a_i$ and the radial velocity is $v/H_0r=
 \dot x/x$.  The peculiar velocity is $v_{\rm pec}/H_0r =
v/H_0r -H/H_0$, where $H$ is the Hubble constant at time $\tau$. 
Thus, for the velocity difference on the caustics, we have 
$\Delta\dot x_3/x_\perp=\Delta v_3/H_0r_\perp$, where $r_\perp$ is the projected physical radius along
the line of sight. Therefore, 
 if we define the amplitude of the velocity field ${\cal A}_v = \Delta v_3/2$, 
we can see that the observable quantity ${\cal A}_v/H_0r_\perp$ 
coincides with the quantity $\Delta\dot x_3/2x_\perp$ we have in space $X$. Fig. 1
shows the profile of the amplitude of the velocity field vs. the local
scale factor  $x_\perp$ of the projected physical radius $r_\perp$ 
for different values of $[\Omega_0,\Omega_{\Lambda0}]$ at the present time. We have 
$r_\perp/r_{\rm vir}=(x_\perp/x_{\rm vir}) (r_\perp/r_{\rm vir})_{a=a_i}$ and a monotonically
decreasing $\delta_i(<r_i)$; 
thus we must stretch these curves out according to $\delta_i(<r_i)$ 
in order to obtain the observable profiles. 
Note that $x_{\rm ta}/x_{\rm vir}\lesssim 2^{5/3}$ is almost constant for $\Omega_{\Lambda0}=0$ universes.
In particular, we have $r_{\rm ta}/r_{\rm vir}=(x_{\rm ta}/x_{\rm vir})(r_{\rm ta}/r_{\rm vir})_{a=a_i}
\gtrsim 3$ in the absence of shell crossings. 

Fig. 1 shows that the caustic amplitude depends on $[\Omega_0,\Omega_{\Lambda0}]$, as expected.
However, for fixed $x_\perp$ the dependence is weak: $\partial\ln\dot x/\partial
\ln \Omega_0\sim0.3$. We recover the usual $0.6$ logarithmic dependence, 
when we switch to physical distances $r_\perp$.

\section{ESCAPE VELOCITY AND MASS ESTIMATE}

In redshift space, the infall regions of halos in $N$-body experiments resemble
the characteristic trumpet shapes expected in the spherical infall model 
(e.g. \cite{HaarWeyg93}; \cite{Jing96}). However, the amplitude of
these caustics, namely the difference between the maximum and minimum line-of-sight velocity at 
projected distance $r_\perp$ from the halo
center, usually exceeds the amplitude predicted by the model. 
The larger amplitude occurs because of random motions in the infall region.

In \S 3.1, we derive an expression for the amplitude of the velocity
field which takes random motions into account. 
This expression indeed replaces the prediction of the spherical infall model. 
Moreover, this expression suggests a method of estimating the mass enclosed 
within a distance $r_\perp$ from the halo center, where $r_\perp$ extends well beyond the
region of virial equilibrium. We derive this expression in \S 3.2.
In \S 4, we compare both expressions with $N$-body simulations of dark matter
halos.
    
\subsection{Amplitude of the velocity field}

Consider a dark halo in its center of mass reference frame. 
Consider the velocity anisotropy parameter at position ${\bf r}$ from the halo center
\begin{equation}
\beta({\bf r}) = 1-{{\langle v_t^2\rangle}\over { 2\langle v_r^2\rangle}}
\label {3.1} 
\end{equation}
where $v_r$ is the radial component of the particle
velocity ${\bf v}$, $v^2=v_t^2+v_r^2$, and $\langle\cdot\rangle$ indicates an average over the 
velocities of all particles within the volume $d^3{\bf r}$ centered on position ${\bf r}$. 
Within the virialized region $r<r_\delta$,
where $r_\delta$ is some virial radius (see \S 4.2.1), 
particle motion is mainly random and $\beta\sim 0$.
For $r>r_\delta$, orbits become more radial and $\beta$ increases up to $\sim 0.8$ 
(see \S 4.2.1).  Within these regions, random
motions still contribute significantly to the velocity field, because
of the intrinsic stochastic accretion process in hierarchical clustering.
The non-zero $\langle v_t^2\rangle$ at large radii is responsible 
for the general increase of the amplitude of the caustics relative to pure spherical
infall. 

In order to write down an expression for the amplitude of the velocity field, we note that
the presence of random motions in the outskirts of the halo implies that
encounters can give a particle enough energy to escape the gravitational
field of the halo. 
Thus, if the random motion is significant, we expect that the amplitude of the velocity 
field at a given distance $r$ {\it is} the line-of-sight component of 
the escape velocity, not the smaller 
velocity induced solely by spherically symmetric infall.

The escape velocity is $v_{\rm esc}^2({\bf r})=-2\phi({\bf r})$: the caustic amplitude is
thus a measure of the gravitational potential $\phi({\bf r})$. For the sake of simplicity,
consider a spherical system. We have
\begin{equation}
v_{\rm esc}^2(r) =-2\phi(r) = {{2GM(<r)}\over {r}} + 8\pi G\int_r^\infty \rho(x)xdx 
\label{3.2}
\end{equation}
where $M(<r)$ is the total mass within $r$ and $\rho(r)$ is the system density profile.
Equation (\ref{3.2}) holds regardless of the stability of the system.

In equation (\ref{3.2}) $v_{\rm esc}$ is the full three-dimensional escape velocity. 
Observations of real clusters will provide only the
component $v_{\rm l.o.s.}$ of the escape velocity along the line-of-sight 
at projected distances $r_\perp$.
The velocity field within the infall region is not isotropic, as indicated by 
the large value of the anisotropy parameter, $\beta\gtrsim 0.5$. 
If the tangential component $v_t(r_\perp)$ of the velocity field is
isotropic, we have $\langle v_t^2\rangle = 2 \langle v_{\rm l.o.s.}^2\rangle $,\footnote{We
are assuming that $v_{\rm esc}^2(r)$ is a non-increasing function of $r$. This behavior is valid
for density profiles steeper than $r^{-1}$.} or
\begin{equation}
\langle v_{\rm esc}^2(r_\perp)\rangle = \langle v_{\rm l.o.s.}^2(r_\perp) \rangle
{{3-2\beta(r_\perp)}\over{1-\beta(r_\perp)}}. 
\label {3.3}
\end{equation}
In \S 4.2.2, we will see that equation (\ref{3.3}) indeed describes the amplitude of
the velocity field of dark matter halos out to a few times the virial radius, $r_\delta$. 

\subsection{Mass estimate}

We now argue that the measure of the escape velocity along the line of sight, i.e. the 
measure of the amplitude of the caustics, may provide a method to estimate $M(<r)$. 
In principle, knowledge of $v_{\rm esc}$ readily yields 
an estimate of the mass within $r$ from equation (\ref{3.2}), 
\begin{equation}
2GM(<r)=-r^2{{dv_{\rm esc}^2}\over{dr}}.
\label {3.4} 
\end{equation}
However, $v_{\rm esc}^2$ is the product of two functions (eq. [\ref{3.3}]). 
The anisotropy parameter $\beta(r_\perp)$ is unknown, but suppose we can model it.
The only measurable function $v_{\rm l.o.s.}^2(r_\perp)$ 
is likely to be very noisy. Thus, differentiation of $v_{\rm esc}^2$ 
is not practical.

We thus consider an alternative approach.
Consider a shell with mass $dm = 4\pi \rho r^2dr$. Assuming $\langle v_{\rm esc}^2\rangle =
-2\phi(r)$, we may write
\begin{equation}
dm = -2\pi \langle v_{\rm esc}^2\rangle { {\rho(r)} r^2\over {\phi(r)}} dr. 
\label {3.5}
\end{equation}
The mass  surrounding the halo is
\begin{equation}
GM(<r) - GM(<r_\delta) =  
\int_{r_\delta}^r \langle v_{\rm esc}^2(x)\rangle \widetilde{\cal F}(x)dx 
\label {3.6}
\end{equation}
where we introduce the filling function $\widetilde{\cal F}(r)=-2\pi G \rho(r)r^2/\phi(r)$.
Here, we integrate the noisy function rather than differentiating it.

We now replace, in equation (\ref{3.6}), the three-dimensional distance $r$ with the projected distance 
$r_\perp$, and the three-dimensional escape velocity 
$v_{\rm esc}$ with the measurable $v_{\rm l.o.s.}$. The second replacement implies the 
introduction of the velocity field anisotropy parameter $\beta$.  Equation (\ref{3.6}) becomes
\begin{equation}
GM(<r_\perp) - GM(<r_\delta) =
\int_{r_\delta}^{r_\perp} \langle v_{\rm l.o.s.}^2(x)\rangle {\cal F}(x)dx 
\label {3.7}
\end{equation}
where
\begin{equation}
 {\cal F}(x) = \widetilde{\cal F}(x) {{3-2\beta(x)}\over{1-\beta(x)}}.
\label{3.8}
\end{equation}

Knowledge of the filling function ${\cal F}$ allows the estimation of the system mass 
for any $r_\perp\in(0,\infty)$. However, we need to know ${\cal F}$ precisely only 
for $r_\perp<r_\delta$. For $r_\perp>r_\delta$ we may consider ${\cal F}$ 
as a free function, generally slowly
varying when $r_\perp/r_\delta\in(1,3)$, where we expect to apply equation (\ref{3.7}).
In fact, the assumption of a spherically symmetric $\rho(r)$ leading to equation (\ref{3.7}) is a very 
crude approximation at best. Thus, the complete expression in equation 
(\ref{3.8}) for ${\cal F}$ might not be 
a robust representation of the filling function. On the other hand, we will see in \S 4.2.3 that 
equation (\ref{3.7}) still holds for reasonable models of ${\cal F}$. 
In practice, we can estimate the mass within $r_\delta$ by applying the virial theorem or by assuming 
hydrostatic
equilibrium of the X-ray emitting gas. Outside $r_\delta$ these methods break down, and we 
can use equation (\ref{3.7}).

Reasonable assumptions about the behavior of
$\rho$ and $\phi$ at large radii provide a model for $\widetilde{\cal F}$. For example, 
if we neglected the integral in equation (\ref{3.2}), we would have 
$\rho\propto r^{-3}$, $\phi\propto r^{-1}$, and $\widetilde{\cal F}= {\rm const}$. 
A universal density profile suggested by Navarro, Frenk \& White (1995) 
for dark matter halos is 
\begin{equation}
\rho(r) = {{\rho_0r_s^3}\over {r(r+r_s)^2}} 
\label {3.9}
\end{equation}
where $r_s$ is some scale length. This profile yields 
\begin{equation}
\widetilde{\cal F}(r) = {{r^2}\over {(r+r_s)^2}} {{1}\over {2\ln(1+r/r_s)}}. 
\label {3.10}
\end{equation}  
This profile is a good fit for $r\lesssim r_\delta$ only. Gravitational lensing observations 
show that this model
reproduces real cluster profiles within $1$ Mpc\footnote{All distances in this paper 
are for $h=0.5$.} (Tormen, Bouchet, \& White 1996). 
However, the model may be a poor approximation in the cluster outskirts.

To obtain ${\cal F}$, we can assume $\beta(r)\sim {\rm const}\sim 0.6-0.8$.
Despite this crude approximation to ${\cal F}$, 
we now show that, for suitable choices of ${\cal F}$, 
the method outlined here can lead to remarkably good mass estimates for $N$-body systems.
 
\section{$N$-BODY SIMULATIONS}

We now use $N$-body simulations to illustrate how random motions in the outskirts of dark
halos invalidate application of the spherical infall model to estimate $\Omega_0$.
We also show that measurement of the amplitude of the velocity field {\it does} provide a 
method of obtaining the enclosed mass with better than $30\%$ accuracy for massive halos.
 
\subsection{Simulation Parameters}

As typical hierarchical clustering scenarios, we consider three Cold Dark Matter (CDM) universes with 
$[\Omega_0,\Omega_{\Lambda0}]=[1.0,0.0],[0.2,0.0],[0.2,0.8]$, and $h=0.5$. 
Initial conditions are generated with the COSMOS package developed by E. \cite{Bertschinger95}
which perturbs initial particle positions and velocities from a grid 
according to the Zel'dovich approximation.
We produce a realization of the Gaussian random field of the initial density perturbations with the
\cite{Bardeen86} power spectrum containing the transfer function for 
adiabatic fluctuations and negligible baryon density. 
We normalize the power spectrum with $\sigma_8=1.0$ at the present time, where
$\sigma_8$ is the rms matter density fluctuation in spheres of radius $16$ Mpc. 
This value of $\sigma_8$ is between the values required to fit the observed abundance of local
clusters in critical and flat low-density CDM universes (\cite{White93}). 
We simulate a 50$^3$ Mpc$^3$ periodic volume with 64$^3$ particles. 

The number density of Abell clusters with richness $R\ge 1$ is 
$\sim 10^{-6}$ Mpc$^{-3}$ (e.g. \cite{Scara91}); thus we need constrained
initial conditions to obtain a rich cluster within the simulation box.
We generate constrained random fields with the algorithm of 
\cite{Hoffman91} in the implementation of \cite{Weyg96}. 
We are interested only in the presence of a rich cluster at the center of the box. Thus, we 
specify only the height $\delta_p$ of the local density maximum. Specifically, we choose
$\delta_p=3\sigma(R_G)$, where $\sigma^2(R_G)$ is the variance of the density field smoothed
with a Gaussian filter of radius $R_G=4$ Mpc.

We integrate the equations of motion with a particle-mesh code (\cite{Hockney81},
\cite{Efstathiou85}) using 128$^3$ mesh points, a seven-point 
finite-difference approximation to the Laplacian, 
a cloud-in-cell density assignment, a leap-frog integrator, 
and an energy conserving scheme to compute the force. 
CDM universes have an effective spectral index $n\sim -1$ on cluster scales. Thus,
we use the scale factor $a$ as the time variable (\cite{Efstathiou85}). 
We also use this variable for the open models, though this choice is actually appropriate
only for flat universes. Simulations obtained with different 
integration variables $a^\alpha$ with $\alpha \in [0.5, 1.5]$ do not yield 
appreciably different results.  
Simulations run from $a\sim 0.02$ to the present time $a=1$ with $\sim 700$ timesteps.
The mesh and the poor momentum conservation of our code 
produce force anisotropies which prevent the Layzer-Irvine
cosmic energy equation from being satisfied to better than $\sim 10$\% over the entire simulation, 
even for timesteps smaller than the timestep we used. 
However, the computed force follows the Newtonian value accurately 
for distances roughly twice the mesh cell size, $\sim 0.4$ Mpc in our simulations. 

Our simulations have two main shortcomings: (1) poor 
spatial resolution and integration accuracy, and 
(2) small box size. Despite the first problem, we will see in the next subsection that
for distances larger than the cell size, our halos roughly reproduce the 
density and velocity fields obtained in 
simulations with larger dynamical range and higher accuracy (\cite{Tormb}). 
At any rate, we are currently running another set of simulations with the AP$^3$M code 
(\cite{Couchman91}) 
made available by Hugh Couchman and collaborators (\cite{Couchman95}) and we
will report on these results in a forthcoming paper.
The second problem, namely the small box size, implies less tractable errors. 

First, periodic boundary conditions produce artificial tidal forces from replicas of structures
within the simulation box. The constrained random field clearly aggravates this problem. 
Artificial tidal forces erroneously increase the random motions in the outskirts of halos; 
they thus artificially worsen the prediction of the spherical infall model.
\cite{Gelbb} suggest that box sizes $L> 50$ Mpc are necessary 
to provide reliable results on clustering. 
However, in $L\sim 50$ Mpc box simulations, individual halo dynamics is almost unaffected because the
artificial tidal forces are sufficiently suppressed (\cite{Gelba}). 
Thus, we expect tidal fields to be a minor problem in our models.

Secondly, we omit contributions from Fourier components of the primordial density perturbation field 
with wavelengths larger than the box size. 
Thus, we underestimate both the amplitude of clustering and 
the amplitude of the peculiar velocities. Corrections to the evolved state of the simulation 
(\cite{Torma}) or correction during its evolution (\cite{Cole96}) alleviate the problem. 
The missing clustering would probably increase the merger rate and therefore
the random motion in the outskirts of halos, thus worsening the predictive power of the
spherical infall model. The velocity field is affected mainly in the bulk
velocity. Random motions increase by only a few percent when one increases the size of the box 
from 100 to 800 Mpc (\cite{Torma}). 

In summary, we increase the random motions 
because of poor spatial resolution and integration accuracy, but we decrease it because of missing
large-scale power. We plan further work to investigate how these problems affect our results.

\subsection{Results}

We now describe the results of our simulations. First, we consider the density and velocity
fields of the most massive halos (\S 4.2.1). In \S 4.2.2, we examine the velocity field within the
infall regions. We compare the simulations with the mass estimation method in \S 4.2.3. 

\subsubsection{Halo Identification}

We identify halos using a generalization of the friends-of-friends
algorithm (\cite{Barnes87}) with a linking length equal to 0.1 Mpc $\sim 0.13$
times the interparticle distance, and a critical number of neighbor particles
$N_c=10$ ($N_c=1$ yields the classical friends-of-friends algorithm). These parameters assure
a halo overdensity $\delta\sim 10^2-10^3$ with respect to the background. 
However, this point is of secondary importance; we are interested only in the
determination of the center of mass of the halo in order to study the velocity field of the outer
regions. Linking lengths between 0.08 and 0.16 Mpc move the center of mass of the 
most massive halos less than 1\%. The center of mass velocity suffers larger oscillations. However,
to suppress this problem 
we redefine the cluster velocity from all the particles within the virial radius $r_\delta$ 
defined below. 

In the outskirts of halos, halo asphericities have little effect on particle velocities, because 
equipotential surfaces are rounder than equidensity surfaces. For example, van Haarlem et
al. (1993) 
find that at distances larger than a few megaparsecs from the halo center, 
the force difference between a
spherical distribution and a significantly ellipsoidal distribution is $\lesssim 15$\%. 
These authors also show that asphericities can blur the caustics expected in the
infall model. However, the disagreement between the spherical infall model and 
the simulations is more severe than the departures induced by asphericity 
(see \cite{HaarWeyg93}). 

We intend to compare the amplitude of the velocity field with the gravitational potential.
Thus, we can assume spherically symmetry, even though halos are not spherically symmetric.
Around the halo center of mass we define spherical shells at equally spaced logarithmic intervals of
distance $r$ from the center. Spherical shells containing the same number of particles
yield the same results.  

In order to compare our simulations with the spherical infall model, we need to determine
the region where the model is valid. Consider the halo density at virialization $\rho_h =
\rho_m/\psi^3(\eta)$, where 
$\rho_m$ is the halo density at maximum expansion, $\psi(\eta)$ is given by equation (\ref{2.9}), and
$\eta= 2\Omega_{\Lambda0}x_{\rm max}^3(\delta_i^c)/\Omega_0(1+\delta_i^c)= \Lambda/4\pi G\rho_m$. 
We obtain 
\begin{equation}
{{\rho_h}\over {\rho(a)} } = {{2\Omega_{\Lambda0}}\over {\Omega_0}} {{a^3}\over {\eta\psi^3(\eta)}} 
\label {4.1}
\end{equation}
where $\rho(a)$ is the mean density of the universe at time $a$. Note that in
$\Omega_{\Lambda0}=0$ universes $2\Omega_{\Lambda0}/\eta=\pi^2/\tau_{\rm coll}^2$, and
equation (\ref{4.1}) reduces to 
\begin{equation}
{{\rho_h}\over {\rho(a)} } = {{8\pi^2}\over {\Omega_0}} {{a^3}\over {\tau_{\rm coll}^2}}.
\label {4.2} 
\end{equation}
In particular, for the Einstein-de Sitter universe the overdensity of the
virialized halo scales as $(1+z_f)^3$, where $z_f$ is the formation redshift of 
the halo (\cite{White96}).
With equation (\ref{4.1}) we can define the radius of the halo $r_\delta$ 
when the average overdensity $\delta(r)$ within the radius $r$ satisfies
$1+\delta=\rho_h/\rho(a)$.
Thus, we expect the spherical infall model to be valid when $r>r_\delta$.

We now consider the halo density and velocity fields. We consider the most massive halos in the
simulations at the present time 
$a=1$, when they are far enough in time from major mergers which formed the final halos.
Strongly unrelaxed states affect the density and velocity fields of the halos (\cite{Tormb})
but have little
effect on the predictive power of the escape velocity as we show in the next subsection. 

Here we examine the halo formation in the three investigated cosmological scenarios on the
basis of simulations starting with the same seed for the random number generator. Thus, we obtain
roughly the same evolutionary history for the three halos. We ran another set of simulations with
different seeds for the random numbers. These simulations provide similar results.
Note that we use a fixed number
of particles within the simulation box; thus, the halo in the flat universe with a final  
mass $\sim 7\cdot 10^{14} M_{\sun}$ is approximately five
times more massive than the most massive halos in the open universes. 

At $a=1$, the largest halos
contain $\gtrsim 20,000$ particles within $r_\delta \sim 2$ Mpc, typically. 
The first row of Fig. 2 shows the circular velocity profile $v_{\rm circ}(r)=[GM(<r)/r]^{1/2}$. 
We compare the simulated profiles with the profile
\begin{equation}
v_{\rm circ}^2 = 4\pi G\rho_0r_s^3 \left[{{1}\over {r}}\ln\left(1+{{r}\over {r_s}}\right) 
-{{1}\over {r+r_s}}\right] 
\label {4.3}
\end{equation}
derived from the Navarro et al. (1995) 
density profile (eq. [\ref{3.9}]). 
Fig. 2 shows the best fits. The agreement is acceptable, even though we fit the profile up to
$r=5r_\delta$ instead of $r\sim r_\delta$ for which equation (\ref{4.3}) is expected to hold
(\cite{Navarro95}). If we fit only the range $[0.1,1]r_\delta$ the agreement improves
slightly, though our poor resolution at small radii is more apparent.
  
The second row of Fig. 2 shows the 
anisotropy parameter $\beta(r)= 1-\langle v_t^2\rangle/2\langle v_r^2\rangle$ (eq. [\ref{3.1}]). 
When $r\gtrsim r_\delta$, $\beta$ increases from $\sim 0.5$ to $\sim 0.6-0.8$, indicating
predominantly radial motion. At $r\sim 5r_\delta$, $\beta$ drops to $\sim 0.2-0.3$ because of
 the presence of another halo. Note that in the flat universe, random motions dominate the velocity
field more strongly
than in the open universes at large $r$. This effect is ${\it not}$ the result of the different
cosmology, but of the different local dynamics and of the larger mass of the halo, which implies
a larger accretion rate (see e.g. \cite{Lacey94}; \cite{Manrique96}). 
For comparison, Fig. 2 shows $\beta(r)$ for a $\sim 2\cdot 10^{14} M_{\sun}$ halo
within the flat universe simulation box (dashed line). The frequency of radial orbits is
larger than for the more massive halo.

Our results agree acceptably with those obtained by Tormen et al. (1996) 
with higher resolution simulations of a $P(k)\propto k^{-1}$ flat universe. Note that 
we reach only a resolution of $\sim
0.1 r_\delta$, compared with the $\sim 0.01 r_\delta$ resolution of Tormen et al. (1996). 

\subsubsection{Velocity Field within the Infall Regions}

We now consider the prediction of the spherical infall model. 
We project the halo along a random direction
and consider the amplitude of the velocity field $2{\cal A}_v(r_\perp)= v_{\rm max}-v_{\rm min}$ 
where $v_{\rm max}$ and $v_{\rm min}$ are the maximum and minimum line-of-sight velocity 
at projected distance $r_\perp$ from the center of the halo. 
Fig. 3 shows ${\cal A}_v/H_0r_\perp$ in our simulations. 
This quantity should agree with the spherical infall prediction for $r_\perp>r_\delta$ (dashed line).
We compute this prediction using complete three-dimensional information to evaluate
the overdensity and then to derive the expected velocity field.
The model systematically underestimates the amplitude of the velocity field. 
Clearly, estimates of $\Omega_0$ based on spherical infall {\it always} overestimate the actual value.

In contrast, agreement with the escape velocity $v_{\rm esc}^2=-2\phi(r_\perp)$ 
(solid line) is excellent. We compute 
the escape velocity through the discrete version of equation (\ref{3.2}). 
We replace the upper limit of integration with a maximum radius $r_{\rm max}$, usually in the range 
$5-10 r_\delta$. Our results are insensitive to this parameter; $r_{\rm max}<5r_\delta$ 
usually underestimates the true potential. When $r_{\rm max}>10r_\delta$ 
the spherical assumption breaks down severely because of the presence of other halos.
We finally correct for the velocity field anisotropy (eq. [\ref{3.3}]): 
\begin{equation}
{\cal A}_v^2(r_\perp) = - 2\phi(r_\perp){{1-\beta(r_\perp)}\over{3-2\beta(r_\perp)}}. 
\label {4.4}
\end{equation}
The agreement is remarkable. It 
indicates that the gravitational potential, i.e. the local dynamics, determines the
amplitude of the velocity field in the outskirts of halos. 
The spherical infall model is a poor predictor of the velocity field.
The global density of the universe apparently plays no role. 

Fig. 3 also shows that the agreement
holds regardless of the dynamical state of the halos. For comparison, we show the halos at
the present time (lower row), when they are approximately in equilibrium, and 
the halos right after the merging of the two halos of comparable size which formed the
final halos (upper row). The escape velocity, namely the gravitational potential, still follows the 
velocity field correctly.

\subsubsection{Estimate of the Halo Surrounding Mass}

We now apply the method outlined in \S 3 to estimate the mass surrounding the
virialized region. We suppose that we know  the virial radius
$r_\delta$, the particle velocities along the line of sight, and
the mass within $r_\delta$, $M(<r_\delta)$. For real systems, we can estimate 
$M(<r_\delta)$ with X-ray methods or with the virial theorem. In future work we plan to investigate
how the uncertainty in the virial mass affects the estimate of the mass within the
infall regions. 

We use equation (\ref{3.7}), assuming 
two different filling functions ${\cal F}$, namely ${\cal F}_1=1/2$, and 
${\cal F}_2 = [\ln(1+\alpha r_\perp/r_\delta)]^{-1}$.
${\cal F}_1$ assumes that the entire gravitational potential originates within $r_\perp$.
${\cal F}_2$  is equation (\ref{3.10}) with a free parameter $\alpha=r_\delta/r_s \sim 3-10$, typically.
For ${\cal F}_2$, 
we have assumed $r^2/(r+r_s)^2\approx 2(1-\beta)/(3-2\beta)$. This assumption is reasonable
when $1\lesssim r/r_\delta\lesssim 5$ and $\beta \sim 0.6-0.7$. 
Results obtained with $\alpha\in (3,10)$ do not differ appreciably. 

Fig. 4 shows the ratio between the estimated mass within the radius $r$, 
$M_{\rm est}(<r)$ and the actual mass 
$M(<r)$. Equation (\ref{3.7}) is apparently a good mass estimator, at least
up to $r\sim 5r_\delta$, where we start observing the velocity field of another halo.
For $r< 5r_\delta$, the difference between the estimated and the actual mass is always $<16\%$, and
is $\sim 10\%$, on average.
Fig. 4 shows both the ``unrelaxed'' (upper row) and ``relaxed'' state (lower row; see Fig. 3).
For the relaxed state the mass estimate improves. However, the difference from the 
unrelaxed state is not large.

If we do not know $M(<r_\delta)$ the agreement between the estimated
mass and the true mass worsens, particularly for unrelaxed states.
Fig. 5 shows the halo mass estimated with equation (\ref{3.7}) when we integrate from $r_\perp=0$.
It is remarkable that ${\cal F}_1$ still works reasonably well for $r_\perp<r_\delta$ when the
clusters are relaxed (lower row), showing
that the filling function ${\cal F}_1$ is more robust than ${\cal F}_2$. We expect this result;
when $r_\perp<r_\delta$ our hypotheses leading to ${\cal F}_2$ break down. 
When the clusters are unrelaxed (upper row) the estimation method introduces large errors. 
These results imply only that both filling functions are inadequate; the velocity field
is still well described by the escape velocity (Fig. 3).

Fig. 6 shows how our mass estimator behaves statistically. For each model, we separate the most 
massive from the least massive halos, regardless of their dynamical state. 
For the flat universe, we consider halos with
$M(<r_\delta)\ge 10^{14} M_{\sun}$ and $10^{13}M_{\sun}\le M(<r_\delta)< 10^{14}
M_{\sun}$. For the open universes, we consider halos with $M(<r_\delta)\ge 2\cdot 10^{13} 
M_{\sun}$ and $10^{12}M_{\sun}\le M(<r_\delta)< 2\cdot 10^{13} M_\odot$.
We thus have roughly the same number of particles within each halo 
for the low or high mass range in both the flat and open universes.
Fig. 6 shows the median of the halo mass profiles; error bars indicate the upper and lower quartile
of all the profiles at each position $r_\perp/r_\delta$. 

The method yields better estimates for more massive halos (upper row).
The agreement for less massive halos worsens because tidal fields caused by 
neighboring larger halos severely perturb the velocity fields of the smaller halos.
However, these small halos contain only a few thousand
particles within $r_\delta\sim $ a few times the cell size. Thus, numerical artifacts may also 
play a significant role. 

Fig. 6 shows that our method of estimating the mass of {\it non-virialized} regions 
yields better than $30$\% statistical accuracy for halos with virial mass 
$M(<r_\delta)\gtrsim 0.5-1.0 \cdot 10^{14}M_{\sun}$. 
For comparison, consider the accuracy of X-ray estimates of the mass of {\it virialized} halos. 
$N$-body/hydrodynamics simulations show that 
the assumption of an isothermal gas in hydrostatic equilibrium yields virial masses 
with better than $20$\% accuracy (\cite{Navarro95}; \cite{Schind96}; \cite{Evrard96}). 
We note however that gravitational lensing
methods, which do not require equilibrium assumptions, may 
disagree with X-ray observations. For example, \cite{Wu96} 
claim that X-ray observations may actually underestimate the mass of real clusters by a 
factor of $\sim 2$. However, gravitational lensing methods suffer systematic errors due to
non-spherical symmetry, projection effects, and substructure (\cite{Miralda95}; 
\cite{Bartelmann95}). On the other hand, 
there are examples where X-ray mass estimates do agree with weak lensing estimates to within the 
error limits (see e.g. \cite{Squires96}; \cite{Squires296}).
We conclude that when applied to $N$-body systems our mass estimation method 
has an accuracy comparable with other widely used methods.

Application of our method to real clusters introduces non-trivial challanges. First, we do not know
how reliably galaxies trace the velocity field in these non-linear regions. Secondly, sampling
effects may introduce large errors, though these effects can be quantified. Even if we could overcome 
these problems, it is not clear
how reliably we can determine the amplitude of the velocity fields in cluster infall regions; the
caustics may not be very apparent (e.g. \cite{Haar93}).
We plan to investigate these issues in future work.

\section{CONCLUSION}

In redshift space, galaxies around clusters should appear within regions with a
characteristic trumpet shape
(\cite{Kaiser87}).  Reg\"os \& Geller (1989) 
suggested applying the spherical infall model to these caustics to constrain the density
of the universe. We show that this method generally overestimates the actual density
parameter $\Omega_0$ because random motions increase the amplitude of the caustics (see
also \cite{Lilje91}; \cite{Zaritsk92}). 

We define the amplitude of the velocity field
as half of the difference between the maximum and minimum line-of-sight velocity at
projected distance $r_\perp$ from the cluster center. We use $N$-body simulations to 
show that the escape velocity (eqs. [\ref{3.2}], and [\ref{4.4}]) describes this amplitude well, with 
$r_\perp$ ranging over two orders of magnitude, from the central region 
of the halo to its infall regions, where particle orbits are mainly radial.
Van Haarlem (1992) 
first noted that the spherical infall method overestimates $\Omega_0$ and pointed out
that mergers and substructures within the infall regions are responsible for the
disagreement. Here we suggest a unifying explanation.

We show that our interpretation of the amplitude of the velocity field
within halo infall regions can be applied to estimate the interior
mass of halos up to a few virial radii $r_\delta$ from the halo center, where the usual
equilibrium assumptions do not  hold. 
This estimation technique works because the local dynamics depends more strongly on the
mass of the halo than on the global properties of the universe. This deduction
agrees with the recent suggestion by White (1996) 
and \cite{Navarro96} that halos have
a universal density profile, basically independent of the cosmology,
 with a characteristic density depending on the formation 
time of the halo (see also our eq. [\ref{4.2}]) which ultimately depends on the halo mass. 
In the literature, the dependence of
the density profile on the halo mass has been neglected, with attention focused
on the dependence on $\Omega_0$ and on the power spectrum $P(k)$. 
The limited overdensity range examined in these numerical studies
explains why $N$-body simulations appear to show a dependence of the halo density profile
on the underlying cosmology (e.g. \cite{Crone94}).

If we know the mass $M(<r_\delta)$ within the virial radius $r_\delta$ our mass estimation 
method can estimate the mass up to several $r_\delta$ with a $\lesssim 30\%$ uncertainty
on average, at least for halos with mass $M(<r_\delta)\gtrsim 0.5-1.0\cdot 10^{14}M_{\sun}$.
If cluster galaxies are unbiased tracers of the gravitational potential, we can
estimate the mass within the outskirts of observed systems on the basis of redshift data alone. 
However, we must investigate how sampling affects the result, and  
we must explore how accurately we need to know the mass within the virial radius.
Last but not least, we must find a reliable method of determining the caustics
and extracting the escape velocity function $v_{\rm esc}(r_\perp)$. This problem
is not trivial because the caustics may not be obvious (e.g. \cite{Haar93}), although
they do appear in some cases (see e.g. A3266 in \cite{Quint96}, and the Coma cluster 
in the 15R Survey of \cite{Geller97}).  
On the theoretical side, we need a reasonable assumption for the filling function ${\cal F}$ that 
we can calibrate with $N$-body simulations. It is reassuring that the simplest assumption we
investigate here, ${\cal F}={\rm const}$, works reasonably well.

We plan to pursue this approach by acquiring 
dense redshift samples in the infall regions of nearby rich galaxy clusters in order
to estimate masses on scales $\lesssim 10$ Mpc, 
where linear theory breaks down and where galaxy systems are not yet in virial equilibrium.

\acknowledgments{We thank Ed Bertschinger for developing the COSMICS package, the cosmological initial
condition generator, and for making it available
to the scientific community. The COSMICS package is funded by the NSF grant AST-9318185.
We also thank an anonymous referee for several constructive suggestions which improved the presentation
of our results.
This research is supported in part by NASA grant NAGW-201 and by the Smithsonian Institution.}

\begin{figure}
\plotfiddle{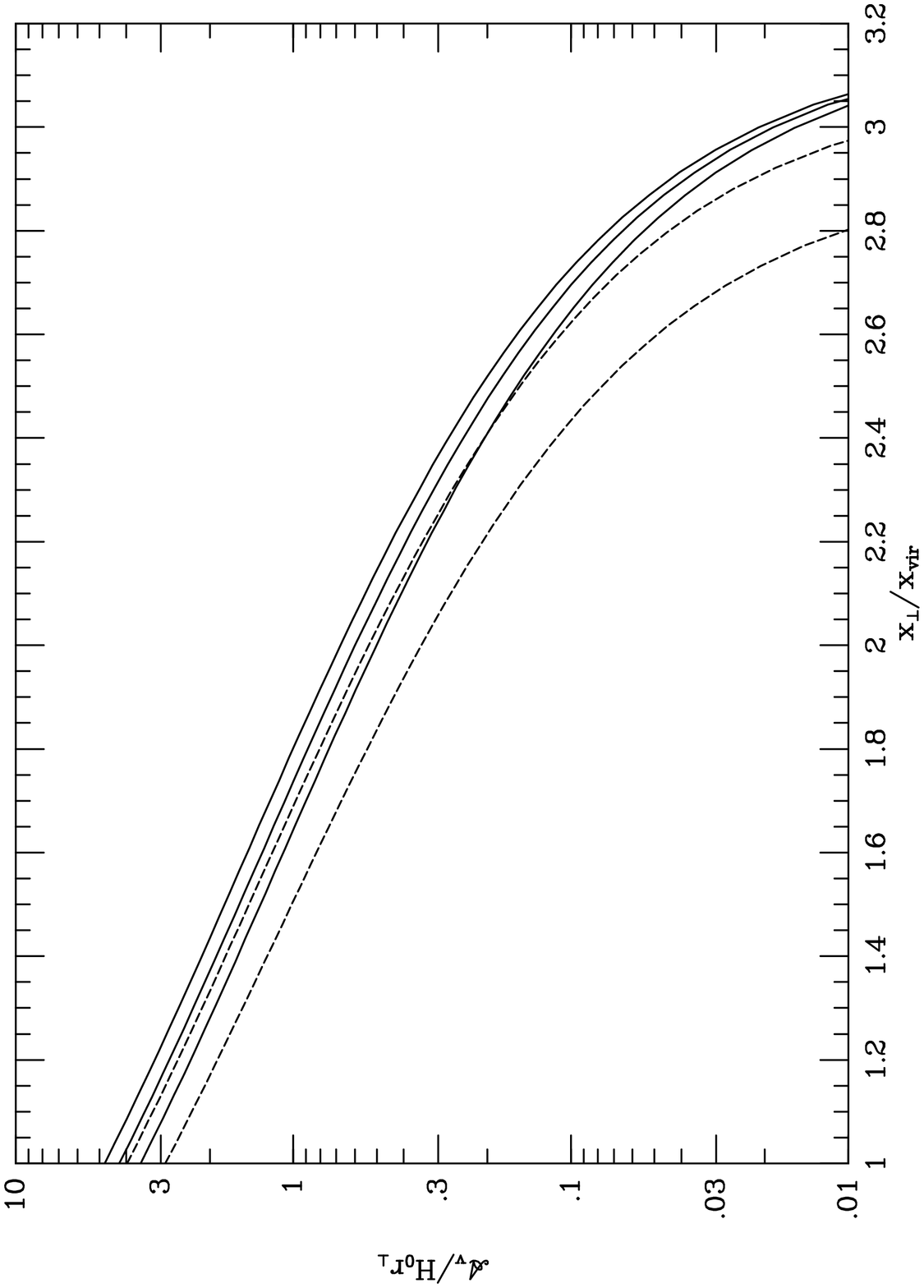}
           {0.4\vsize}              
           {270}                
           {60}                 
           {60}                 
           {-270}               
           {340}                
\caption{Amplitude ${\cal A}_v$ of the velocity field within the infall region
of a spherical perturbation in an otherwise uniform universe at the present time.
The amplitude is in units
of the projected physical distance $r_\perp$ from the center of the perturbation; $x_\perp$ is the
component perpendicular to the line of sight of the local scale factor $x$.
From top to bottom, solid lines are for
$[\Omega_0,\Omega_{\Lambda0}] = [1.0, 0.0], [0.5, 0.0], [0.1, 0.0]$.
Upper dashed line is for $[\Omega_0,\Omega_{\Lambda0}] = [0.5, 0.5]$; lower
 dashed line is for $[\Omega_0,\Omega_{\Lambda0}] = [0.1, 0.9]$.}
\end{figure}

\begin{figure} 
\plotfiddle{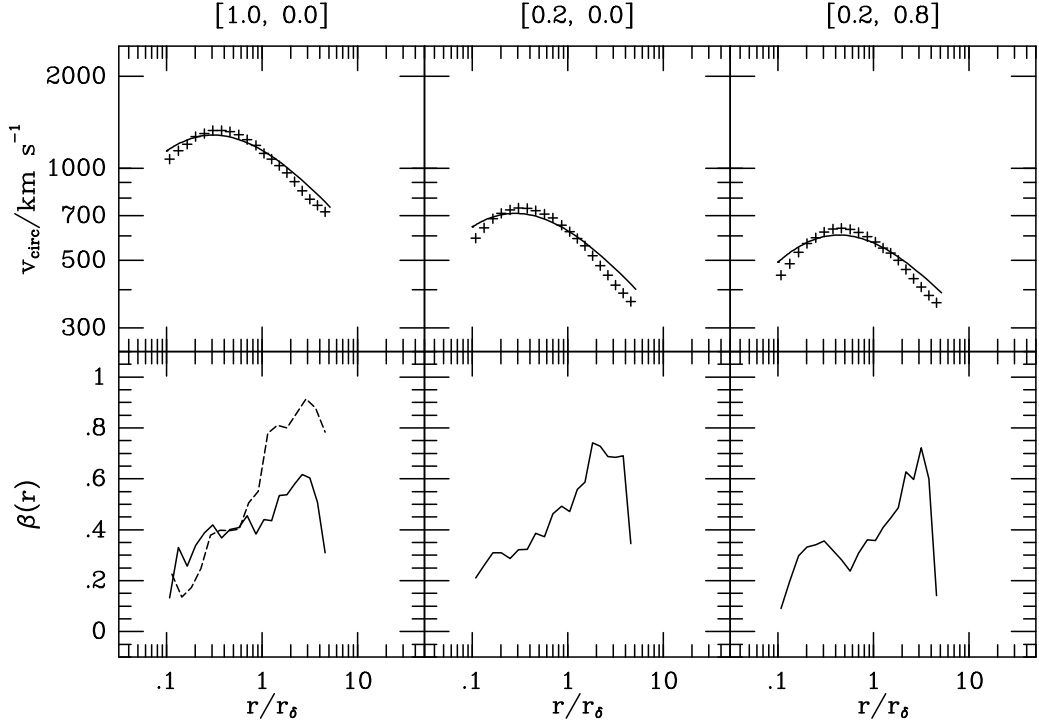}
           {0.4\vsize}              
           {270}                
           {60}                 
           {60}                 
           {-270}               
           {340}                
\caption{Density and velocity fields of the most massive halo in each
$N$-body model. Upper row shows  the circular velocity $v_{\rm circ} = [GM(<r)/r]^{1/2}$. Solid
lines are the best fits to equation (\ref{4.3}). Lower row shows the velocity field
anisotropy parameter $\beta(r)$ (eq. [\ref{3.1}]). The values of $[\Omega_0,\Omega_{\Lambda0}]$
are shown over each column. The dashed line in the $[\Omega_0,\Omega_{\Lambda0}]=[1.0,0.0]$
model is for a halo of mass $M\sim 2\cdot 10^{14} M_{\sun}$, roughly the
most massive halo in the open models.}
\end{figure}

\begin{figure}
\plotfiddle{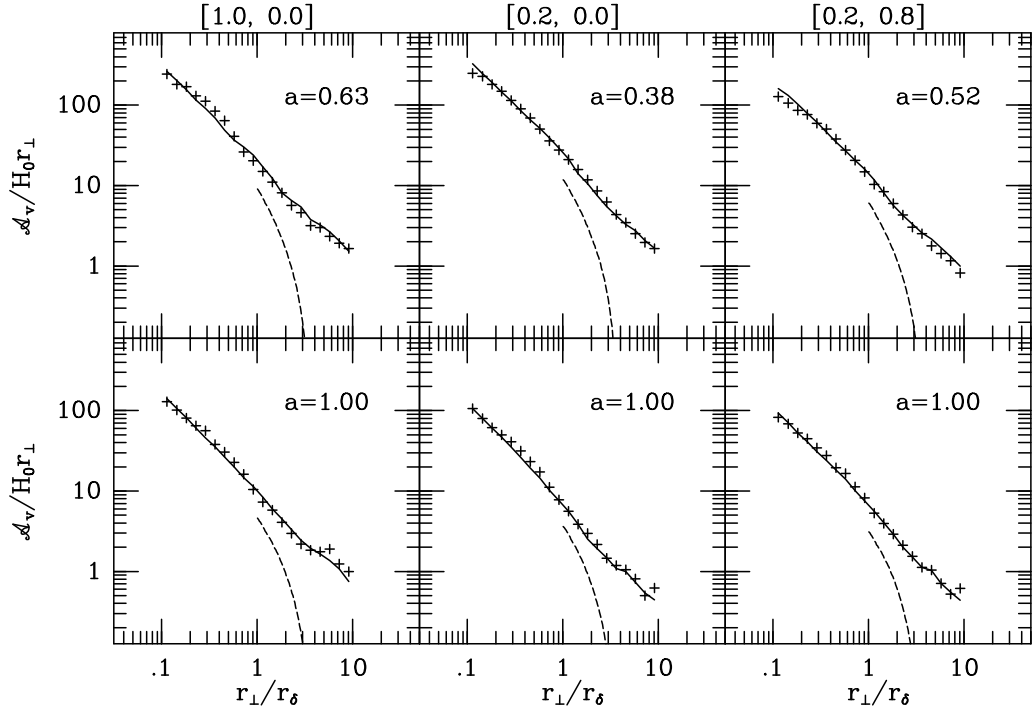}
           {0.4\vsize}              
           {270}                
           {60}                 
           {60}                 
           {-270}               
           {340}                
\caption{Amplitude of the velocity field within the infall region
of the most massive halo in each $N$-body model. Solid lines are the escape velocities
computed with equation (\ref{3.2}) corrected for the anisotropy parameter with equation (\ref{3.3}).
Dashed lines are the spherical infall predictions which hold only for $r_\perp>r_\delta$.
Lower row shows the halos at the present
time $a=1$, when the halos are roughly in equilibrium. Upper row shows the
halos at earlier times, right after the major mergers of the two smaller halos which
formed the final halos.}
\end{figure}

\begin{figure}
\plotfiddle{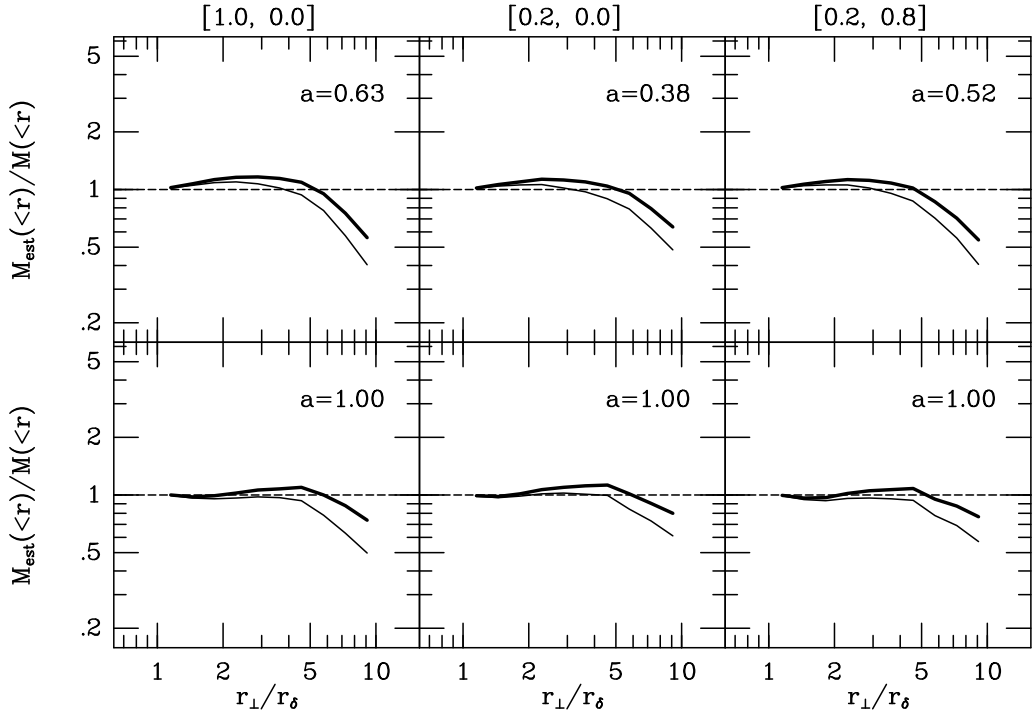}
           {0.4\vsize}              
           {270}                
           {60}                 
           {60}                 
           {-270}               
           {340}                
\caption{Ratio of the interior mass $M_{\rm est}(<r)$ 
estimated with equation (\ref{3.7}) and the true interior mass $M(<r)$ of the most massive
halo in each $N$-body model. We assume that we know $r_\delta$ and the virial mass $M(<r_\delta)$.
Bold lines are for the filling function ${\cal F}_1=1/2$. Solid lines are for ${\cal F}_2
= [\ln(1+7r_\perp/r_\delta)]^{-1}$. Times are as in Fig. 3.}
\end{figure}

\begin{figure}
\plotfiddle{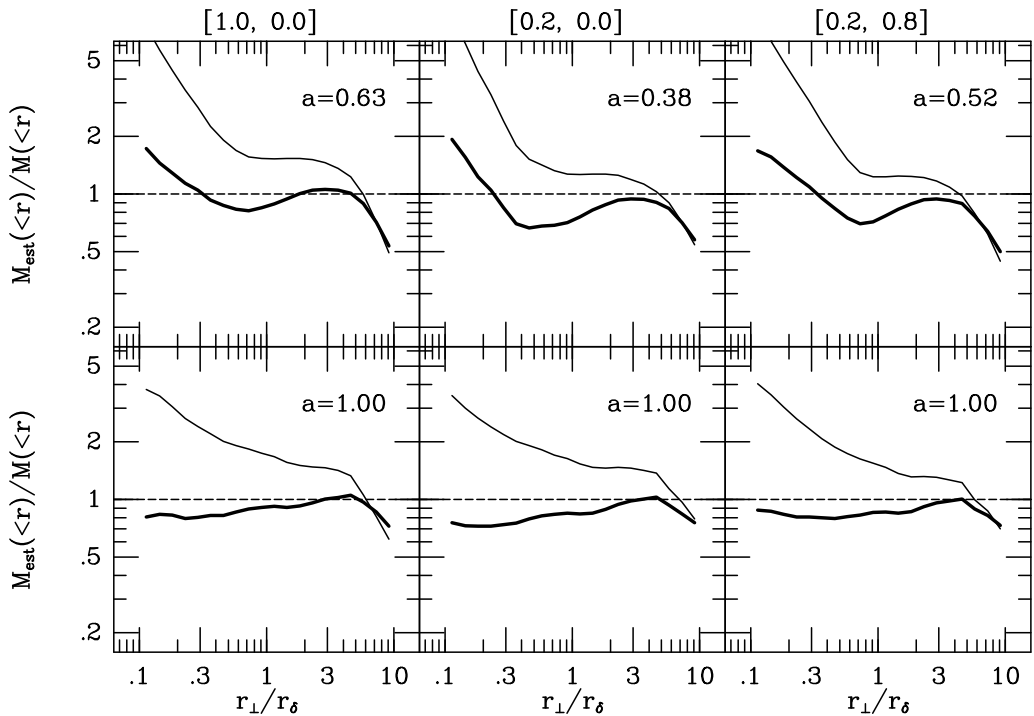}
           {0.4\vsize}              
           {270}                
           {60}                 
           {60}                 
           {-270}               
           {340}                
\caption{Same as Fig. 4, but we now assume that we do not know the virial 
mass $M(<r_\delta)$ and we integrate equation (\ref{3.7}) from $r_\perp=0$.}
\end{figure}

\begin{figure}
\figurenum{6a}
\plotfiddle{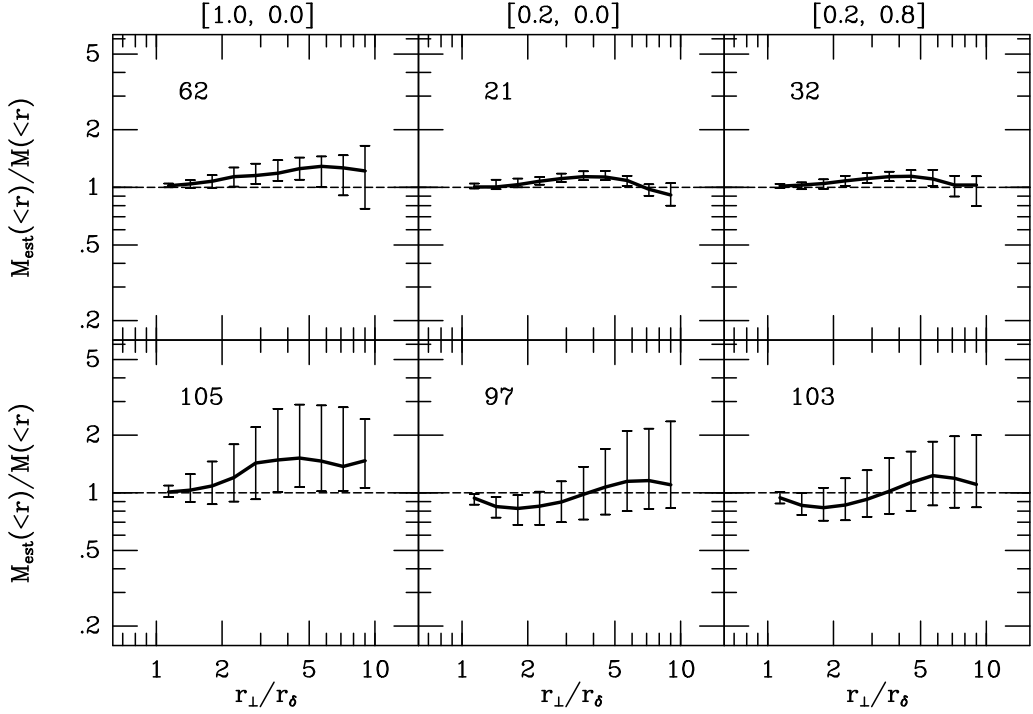}
           {0.4\vsize}              
           {270}                
           {60}                 
           {60}                 
           {-270}               
           {340}                
\caption{Median mass profiles of halo samples in each $N$-body model.
Masses are estimated with equation (\ref{3.7}), assuming we know each $M(<r_\delta)$.
Upper rows show the most massive halos: $M(<r_\delta)\ge 10^{14} M_{\sun}$ for the flat
model, and $M(<r_\delta)\ge 2\cdot 10^{13}M_{\sun}$ for the open models.
Lower rows show the least massive halos: $10^{13}M_{\sun}\le
M(<r_\delta)< 10^{14} M_{\sun}$ for the flat model, and
$10^{12}M_{\sun}< M(<r_\delta)\le 2\cdot 10^{13} M_\odot$ for the open
models. Numbers of halos in each sample are shown. Error bars indicate upper and lower
quartiles at each projected distance $r_\perp$. (a) Filling function ${\cal F}_1=1/2$;
(b) filling function ${\cal F}_2= [\ln(1+7r_\perp/r_\delta)]^{-1}$.}
\end{figure}

\begin{figure}
\figurenum{6b}
\plotfiddle{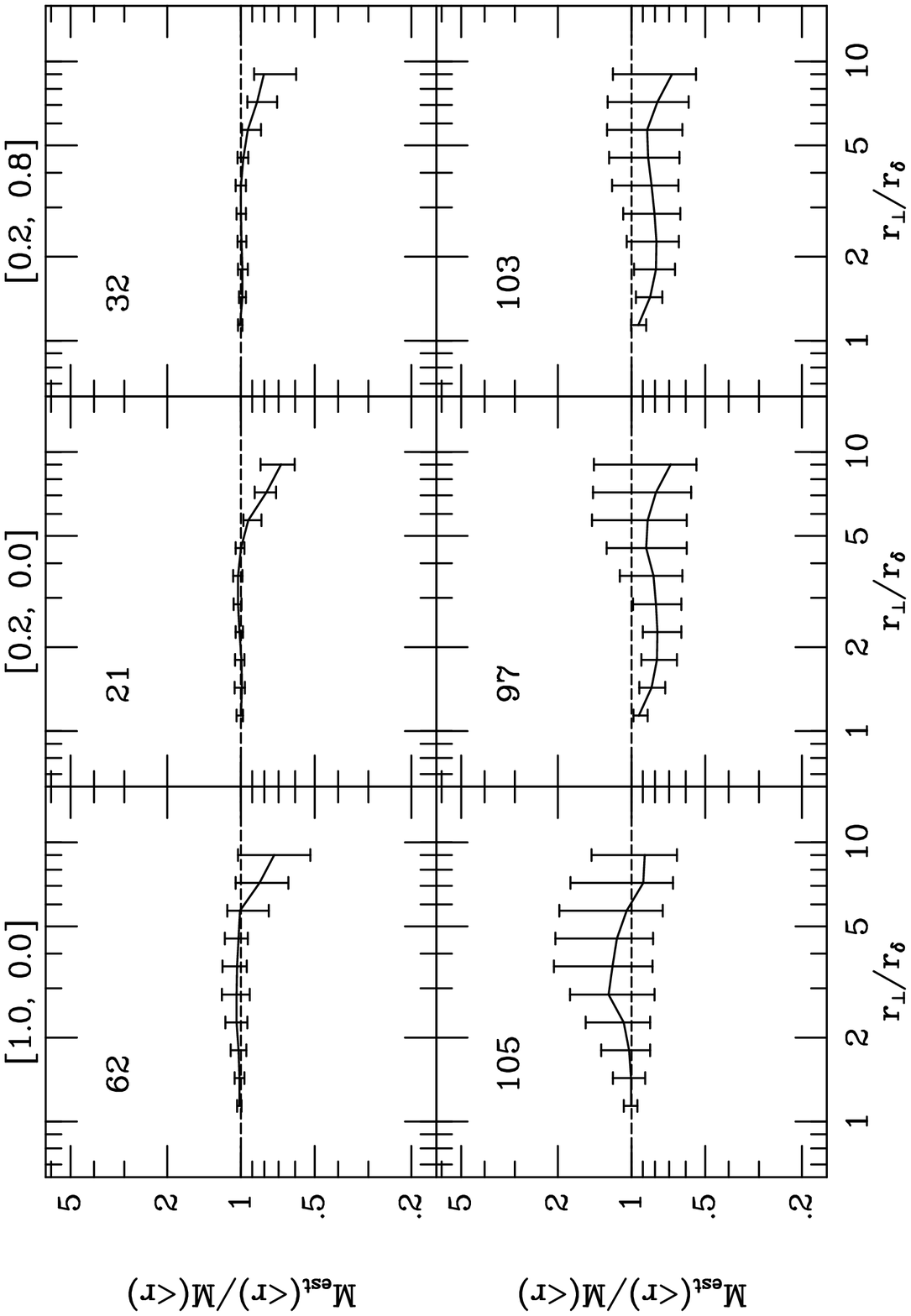}
           {0.4\vsize}              
           {270}                
           {60}                 
           {60}                 
           {-270}               
           {340}                
\caption{}
\end{figure}

\end{document}